\documentstyle[12pt]{article}
\begin{document}

\begin{titlepage}
\rightline{May 1995}
\rightline{UM-P-95/49}
\rightline{RCHEP-95/14}
\vskip 2cm
\centerline{\large \bf Neutrino physics and the mirror world:}
\vskip .2cm
\centerline{\bf How exact parity symmetry explains
the solar neutrino deficit,}
\vskip .1cm
\centerline{\bf
the atmospheric neutrino anomaly and the LSND experiment}
\vskip 1.1cm
\centerline{R. Foot and R. R. Volkas}
\vskip .7cm
\centerline{{\it Research Centre for High Energy Physics}}
\centerline{{\it School of Physics}}
\centerline{{\it University of Melbourne}}
\centerline{{\it Parkville 3052 Australia}}
\vskip 2cm

\centerline{Abstract}
\vskip 1cm
\noindent
Evidence for $\bar \nu_{\mu} \rightarrow \bar \nu_e$
oscillations has been reported at LAMPF
using the LSND detector.
Further evidence for neutrino mixing comes from
the solar neutrino deficit and the atmospheric neutrino
anomaly.
All of these anomalies require new physics.
We show that all of these anomalies can be explained if
the standard model is enlarged so that an unbroken parity
symmetry can be defined. This explanation holds
independently of the actual model for neutrino masses.
Thus, we argue that parity
symmetry is not only a beautiful candidate for a symmetry
beyond the standard model, but it can also explain the
known neutrino physics anomalies.

\end{titlepage}
\noindent
{\large \bf I Introduction}
\vskip 0.7cm
\noindent
Recently, the LSND Collaboration has found evidence for
$\bar \nu_{\mu} \rightarrow \bar \nu_e$ oscillations [1].
If the anomaly in this experiment is interpreted as neutrino
oscillations, then they obtain the range of parameters
$\Delta m^2 \sim 3 - 0.2 \ eV^2$
and $\sin^2 2\theta \sim 3\times 10^{-2} - 10^{-3}$.
If the interpretation of this experiment is correct then
it will lead to important ramifications for particle physics and
cosmology.

In addition to the direct experimental anomaly discussed above,
and the theoretical argument for non-zero neutrino masses from
the observed electric charges of the known particles
[2], there are two indirect
indications
that the minimal standard model is incomplete.
Firstly, there are the solar neutrino experiments [3].
There are four experiments
which we summarise in table 1, below:
\vskip 0.7cm

\tabskip=0pt \halign to \hsize{
\vrule#\tabskip=0pt plus 1fil\strut&
\hfil#\hfil& \vrule#& \hfil#\hfil& \vrule#& \hfil#\hfil& \vrule#& \hfil#\hfil&
\tabskip=0pt\vrule#\cr
\noalign{\hrule}
&Experiment&&Measurement&&SSM-BP&&SSM-TCL&\cr
\noalign{\hrule}
&$^{37}$Cl[SNU]&&$2.23\pm0.23$&&$8\pm 1$&&$6.4 \pm 1.4$&\cr
&Kamioka $\left({Observed \over SSM-BP}\right)$&&$0.50
\pm 0.04 \pm 0.06$&&$1 \pm 0.14$&&$0.77 \pm 0.17$&\cr
&GALLEX[SNU]&&$79\pm 10 \pm 6$&&$131.5\pm 7$&&$122.5\pm 7$&\cr
&SAGE[SNU]&&$73^{+18+5}_{-16-7}$&&"&&"&\cr
\noalign{\hrule}
}

\vskip 0.8cm
\noindent
{\bf Table 1}  Solar neutrino measurements and
theoretical expectations within the
Standard Solar Model (SSM) of Bahcall and
Pinsonneault [4], SSM-BP, and Turck-Chieze
and Lopes [5], SSM-TCL.
\vskip .5cm
\noindent
Note that the theoretical predictions
for the flux of solar neutrinos
involve a lot of assumptions, and the true theoretical
value may be outside these errors. In particular the  analysis
by Turck-Chieze and Lopes [5] gives theoretical predictions
for the experiments which are different to those
of Bahcall and Pinsonneault [4] but which still seem
to be too high to be consistent
with the data (although it has been argued that the data and
the theory may be in-agreement if one takes into account
all sources of uncertainty [6]). In view of the above, we do
not attempt to propose a particular physics solution which will make
all of the experiments agree with the theoretical prediction of
Bahcall et al [4] (as is sometimes done).
If there is a solar neutrino problem and if
new particle physics is the solution then any new particle physics
which can reduce the number of solar neutrinos by a large fraction
(e.g. 1/2) may be the cause of the apparent disagreement
of theory with data.
One interesting possibility, which we will assume, is that the
deficit of solar neutrinos is due to vacuum neutrino oscillations [7].
In particular, if the electron neutrino is a maximally mixed
combination of two states,
then the number of neutrinos
expected from the sun will be 0.5 that of the standard model (for
a large range of parameters). This would be, in our opinion, an adequate
``explanation'' of the solar neutrino deficit [8].

Another experiment which seems to be in conflict with theory
is the atmospheric neutrino experiment [9]. This experiment measures
the ratio of $\nu_{\mu}/\nu_e$ interactions where the neutrinos
are presumed to originate from cosmic ray
interactions in the atmosphere. These experiments observe a deficit
in the ratio of $\nu_{\mu}/\nu_e$ interactions when the data
is compared with theory. We summarise the situation below:
$$\begin{array}{ll}
Kamiokande&0.60\pm0.07(stat)\pm0.05(sys),\\
IMB&0.55\pm 0.05(stat)\pm 0.10(sys),
\end{array}
\eqno (1)$$
where the data has been normalized to the theoretically expected
ratio [10].
Recently, the Kamiokande group has examined atmospheric
neutrino events with higher energy, $\stackrel{>}{\sim} 1.3$ GeV [11].
Because of the higher energy, they can search for path length
dependence.  They show that their data can be fit
if the muon neutrino is a mixture of two states with the
parameters
$$\Delta m^2 \simeq 10^{-2} eV^2\ {\rm and} \ \sin^2 2\theta
\simeq 1. \eqno (2)$$
This result, together with the earlier Kamiokande and IMB results
strongly suggest that the muon neutrino is, at least approximately,
a maximally mixed combination of two states.

If intergenerational mixing is suppressed as it is in the
quark sector, and also as suggested by the LSND experiment,
then the only way to get vacuum oscillations large enough
to explain the solar neutrino and atmospheric neutrino anomalies
is if there exist additional light neutral particles.
These additional light particles cannot belong to additional
generations of the usual
form, since these types of neutrinos couple to the Z-boson
and are ruled out by LEP experiments. The only remaining
possibilities are
that the additional light neutrinos are either
\newline
\noindent
a) gauge singlets,
\newline
\noindent
b) are members of exotic $SU(2)_L$ multiplets [12], or
\newline
\noindent
c) are members of multiplets of a gauge symmetry [which
is not $SU(3)_c \otimes SU(2)_L \otimes U(1)_Y$].

While gauge singlets might exist, we would expect them to
be heavy, since their masses are not protected by electroweak
symmetry. Similarly, if the additional
neutrinos are members of non-chiral multiplets, gauge symmetry
does not protect their masses, and we would expect them to be
heavy.
Exotic $SU(2)_L$ multiplets may exist, but are
probably unlikely since the charged members of the multiplets
have to be sufficiently heavy to avoid being detected in the
decays of the Z, W gauge bosons. Also,
the additional contributions to the oblique radiative corrections,
which would be expected to be quite large, would have to cancel
in order to reproduce the success of the standard model.
This seems unlikely. Thus, the only remaining possibility
is that the additional light neutrinos are chiral members of
a gauge symmetry not contained in $SU(3)_c \otimes SU(2)_L \otimes
U(1)_Y$. This additional gauge symmetry must be broken (since
if it was unbroken, the additional neutrinos would be massless,
and could not mix with the ordinary neutrinos), and the scale
of symmetry breaking must be less than or not much greater than
the electroweak symmetry breaking scale (since otherwise, the
gauge symmetry would not protect the masses of the additional
neutrinos).
[This observation would rule out simple grand unified
models such as $SO(10)$ as
candidates, since, although they contain additional neutral neutrino
species, $SO(10)$ symmetry breaking to the standard model gauge
symmetry must occur at a scale significantly higher than the
electroweak scale. The usual left-right symmetric model is also
not appropriate since the scale of $SU(2)_R \otimes U(1)_{B-L}$
breaking must also be significantly higher then the electroweak
breaking scale.]

In other words, the assumption that intergenerational mixing
is small, which is supported by the LSND experiment and
the small intergenerational mixing of the quarks,
together with the large vacuum oscillations needed
to explain the solar neutrino deficit and atmospheric neutrino
anomaly, imply the existence of additional neutrino species.
Furthermore, theoretical arguments suggest that these
additional neutrino species should be chiral members of
a gauge symmetry not contained in $SU(3)_c \otimes SU(2)_L \otimes U(1)_Y.$
This additional gauge
symmetry should be broken at a scale less than, or
not much greater than the electroweak symmetry breaking scale.

We believe that the most compelling model which contains additional
neutrino species which are chiral members of
a gauge symmetry not contained in $SU(3)_c \otimes SU(2)_L \otimes
U(1)_Y$,
is the standard model extended to include an exact parity symmetry.
It has been known for a long time, but not widely appreciated,
that an exact parity symmetry can be defined if the particle
content and the gauge symmetry of the standard model is doubled [13].
Not  only does exact parity symmetry demand the existence
of additional light neutrino states,
but more importantly
if neutrinos are massive and parity is
unbroken, then the weak eigenstate will
be maximally mixed combinations of mass eigenstates [14, 15].

This means that for a large allowed range of parameters,
the flux of electron neutrinos from the sun will be {\it predicted } to
be half
that of the standard model, and the number of atmospheric muon
neutrinos will be {\it predicted } to be half that of the standard model.
{\it Both of these predictions are in good agreement with the data,
whereas the standard model is not}.

The outline of this paper is as follows: In section II we
review the exact parity model. In section III we show that
if neutrinos have mass and the ordinary and mirror neutrinos
mix together, then in general the oscillations will be maximal.
We show that this result can explain the solar neutrino deficit,
the atmospheric neutrino anomaly, and is consistent with the LSND
experiment. In section IV we give a naturalness argument, which indicates
that the parameters needed to explain the atmospheric neutrino deficit
and the LSND result together imply that the expected range of
$\delta m^2$ for
solar neutrinos is in the correct range to explain the deficiency.
In section V we illustrate the results of the previous experiments
in a concrete model for neutrino masses. The model we use is the
usual see-saw model extended so that it is parity symmetric.
In section VI we examine a variant of the exact parity model, the
exact C invariant model, which also has a mirror sector. This model
has very similar predictions and is in general very similar
to the exact parity model.
In section VII we discuss the incompatibility of the exact parity
model (and exact C model) with the standard big bang model of cosmology.
In section VIII we conclude with some comments.
\vskip 0.7cm
\noindent
{\large \bf II Exact parity symmetric models}
\vskip 0.5cm
\noindent
In the important paper on possible parity violation in weak
interactions, Lee and Yang [13] not only suggested that parity could
be violated in the weak interactions, but also pointed out
that parity could be retained by enlarging the particle content
to include a mirror sector.
Since that time, a number of authors have returned to that
idea [17, 18, 19].
In these works (with the exception of Ref. [18] as we shall discuss
later), it was thought that the mirror sector could
not interact with ordinary matter, and hence was of only
cosmological [19] and philosophical interest.
Independently of these early works, H. Lew and ourselves [16]
realized that parity could be conserved by enlarging the particle
content to include a mirror sector. In our  paper we wrote down
the Lagrangian for that theory, we showed that it was
sensible and unbroken parity was a possible vacuum of the Higgs
potential. We also observed that the mirror sector could
in fact interact with ordinary matter, and hence the idea is
testable in the Laboratory.

We now review the exact parity symmetric
model [16].
To understand how parity might be conserved, consider
a model which successfully describes present experiments.
In particular, consider the minimal standard model.
This model is described by a Lagrangian ${\cal L}_1$.
This Lagrangian is not invariant under the usual parity
transformation so it seems parity is violated. However,
this Lagrangian may not be complete.
If we add to ${\cal L}_1$ a new Lagrangian ${\cal L}_2$
which is just like ${\cal L}_1$ except that all left-handed
(right-handed) fermions are replaced by new right-handed
(left-handed) fermions which feel new interactions of the same
form and strength, then the theory described by
${\cal L} = {\cal L}_1 + {\cal L}_2$ is invariant under
a parity symmetry (under this symmetry ${\cal L}_1
\leftrightarrow {\cal L}_2$). In addition to these Lagrangian terms,
there may also be terms which mix ordinary with mirror
matter and which are parity invariant.
We label this part of the Lagrangian as ${\cal L}_{int}$.
The terms in ${\cal L}_{int}$ are very important since they lead
to interactions between ordinary and mirror matter, and hence
allow the idea to be experimentally tested in the laboratory.

If we apply the above procedure to the standard model then
${\cal L}_1$ is just the standard model Lagrangian.
We now add the ``mirror matter'' as described above,
so that the total Lagrangian consists of
two parts ${\cal L}_1$ and ${\cal L}_2$.
Then the gauge symmetry of the theory is
$$SU(3)_1 \otimes SU(2)_1 \otimes U(1)_1 \otimes
SU(3)_2 \otimes SU(2)_2 \otimes U(1)_2. \eqno (3)$$
There are two sets of fermions, the ordinary particles
and their mirror images, which transform under the
gauge group of Eq.(3) as
$$\begin{array}{ll}
f_L \sim (1, 2, -1)(1, 1, 0),& F_R \sim (1, 1, 0)(1, 2, -1),\\
e_R \sim (1, 1, -2)(1, 1, 0),& E_L \sim (1, 1, 0)(1, 1, -2),\\
q_L \sim (3, 2, 1/3)(1, 1, 0),&Q_R \sim (1, 1, 0)(3, 2, 1/3),\\
u_R \sim (3, 1, 4/3)(1, 1, 0),&U_L \sim (1, 1, 0)(3, 1, 4/3),\\
d_R \sim (3, 1, -2/3)(1, 1, 0),&D_L \sim (1, 1, 0)(3, 1, -2/3),
\end{array}
\eqno (4)
$$
(with generation index suppressed). The Lagrangian is invariant
under the discrete $Z_2$ parity symmetry defined by
$$\begin{array}{c}
x \rightarrow -x,\  t \rightarrow t,\\
G_1^{\mu} \leftrightarrow G_{2\mu},\
W_1^{\mu} \leftrightarrow W_{2\mu},\ B_1^{\mu}
\leftrightarrow B_{2\mu},\\
f_L \leftrightarrow \gamma_0 F_R,\ e_R \leftrightarrow
\gamma_0 E_L,\ q_L \leftrightarrow \gamma_0 Q_R,\
u_R \leftrightarrow \gamma_0 U_L,\ d_R \leftrightarrow
\gamma_0 D_L, \end{array}
\eqno (5) $$
where $G_1^{\mu}(G_2^{\mu}), W_1^{\mu} (W_2^{\mu}) $
and $B_1^{\mu} (B_2^{\mu})$ are the gauge bosons of
the $SU(3)_1$ $[SU(3)_2]$,
$SU(2)_1 [SU(2)_2], U(1)_1 [U(1)_2]$
gauge forces respectively.
The minimal model contains two Higgs doublets which are
also parity partners:
$$ \phi_1 \sim (1, 2, 1)(1, 1, 0),\  \phi_2 \sim (1, 1, 0)(1, 2, 1).
\eqno (6)$$
Note that, although the parity symmetry is not of the standard form, it
is theoretically a perfectly reasonable candidate for a parity symmetry. It
commutes with the proper
Lorentz group, interchanges x with -x and leads to
a theory described by a Lagrangian which treats
left and right on an equal footing.
Also, by virtue of the CPT theorem, there will
be an unbroken T symmetry
which also connects
the ordinary and mirror particles as partners
(so that CPT has the usual form),

An important feature
which distinguishes this parity conserving theory
from other such theories [20]
is that the parity symmetry is assumed to be
unbroken by the vacuum.
The most general renormalizable Higgs potential
can be written in the form
$$V(\phi_1, \phi_2) = \lambda_1 (\phi_1^{\dagger} \phi_1
+ \phi_2^{\dagger} \phi_2 - 2u^2)^2 + \lambda_2 (
\phi_1^{\dagger} \phi_1 - \phi_2^{\dagger} \phi_2)^2, \eqno (7)$$
where $\lambda_{1,2}$ and $u$ are arbitrary constants. In
the region of parameter space where $\lambda_{1,2} > 0$,
$V(\phi_1, \phi_2)$ is non-negative and is minimized
by the vacuum
$$\langle \phi_1 \rangle = \langle \phi_2 \rangle =
\left( \begin{array}{c}
0\\
u
\end{array}\right).
\eqno (8)$$
The vacuum values of both Higgs fields are exactly the same
and hence parity is not broken by the vacuum in this
theory [21].

If the solar system is dominated by the usual particles,
then the theory agrees with present experiments.
The idea can be tested in the laboratory because it is possible
for the two sectors to interact
with each other via ${\cal L}_{int}$. In the simplest case that we
are considering at the moment (where ${\cal L}_1$ is
the minimal standard
model lagrangian), there are just two possible terms (i.e.
gauge invariant and renormalizable) in ${\cal L}_{int}$.
They are,
\vskip .3cm
\noindent
(1) The Higgs potential terms $\lambda \phi_1^{\dagger} \phi_1
\phi_2^{\dagger} \phi_2$ in Eq.(7) and
\vskip .3cm
\noindent
(2) The gauge boson kinetic mixing term ${\cal L}_{mix} = \delta
F^1_{\mu \nu} F^{2 \mu \nu}$,
where $F^{i}_{\mu \nu} = \partial_{\mu} B^{i}_{\nu} -
\partial_{\nu} B^{i}_{\mu}$ ($i=1,2$).
\vskip .3cm
\noindent
The principal phenomenological effect of the Higgs
potential mixing term in (1) is to modify
(quite significantly)
the interactions of the Higgs boson. This effect will be tested
if or when the Higgs scalar is discovered. The details have been
discussed in Ref.[14]. The principal phenomenological
effect of the kinetic mixing term in (2) is to give small
electric charges to the mirror partners of the ordinary charged
fermions. This effect has also been discussed previously [16, 18, 22].
The experimental bounds on $\delta$ are quite weak (about $10^{-3}$
from searches for minicharged particles [23]), but a new experiment
is underway [24] which will either improve
this bound, or discover a minicharged
fermion. If this fermion has the same mass as one of the known charged
fermions, e.g. the electron mass, then it would give strong experimental
support for the parity conserving model.
Looking for minicharged particles and studying the properties of the Higgs
boson are the only two ways to experimentally
test the minimal parity conserving
model. Thus,
despite the fact that exact parity symmetry necessarily
predicts the existence of new light states, the observed agreement of
present experiments (not including the neutrino anomalies)
with the standard model
can also be viewed as evidence for the parity conserving model.
In fact, given that the Higgs particle is heavier than the LEP bound,
the only way to discover the new physics predicted by
the exact parity model (with existing experiments)
is to search for mini-charged fermions
which have the same masses as the known charged fermions.
The situation changes dramatically if neutrinos have
mass [14, 15] as we shall now discuss.

\vskip 0.7cm
\noindent
{\large \bf III Neutrino mass and the exact parity symmetric model}
\vskip 0.5cm
\noindent
If neutrinos are massive,
then this will provide an important new way for the mirror world to
interact
with the known world [14, 15].
This is important since it will allow the idea that parity
is a exact symmetry of nature to be put to further
experimental test.
If neutrinos are massive, then
${\cal L}_{int}$ can contain neutrino
mass terms which mix the ordinary and mirror matter. (Note
that if electric charge is conserved,
then it is not possible for ${\cal L}_{int}$ to contain
mass terms mixing the charged fermions of ordinary matter
with mirror matter, however neutrinos may be neutral so
such mass terms are possible provided that the neutrinos
have masses).

To see the effect of the mixing of ordinary and mirror matter
consider
the electron neutrino. If there were no mirror matter,
then small intergenerational mixing will imply that
the weak eigenstate electron neutrino will be approximately
a single mass eigenstate, $\nu_e$, with mass $m$. However
if mirror matter exists, then there will be a mirror
electron neutrino $\nu_E$. If neutrinos are Majorana
states, then the most general mass matrix consistent
with parity conservation [Eq.(5)] is
$${\cal L}_{mass} = [\bar \nu_{eL}, (\bar \nu_{ER})^c]
\left(\begin{array}{cc}
m&m'\\
m'&m^*
\end{array}\right)\left[
\begin{array}{c}
(\nu_{eL})^c\\
\nu_{ER}
\end{array}\right] + H.c. \eqno (9)$$
where $m'$ is real (due to parity symmetry).
Observe that the parameter $m$ can be
taken to be real by a choice of phase for $\nu_e$ and $\nu_E$.
Diagonalizing this mass matrix, we easily obtain that
the weak eigenstates $\nu_e, \nu_E$ are each maximally mixed
combinations of mass eigenstates:
$$\nu_{eL} = {(\nu_1^+ + \nu_1^{-})_L \over \sqrt 2},$$
$$(\nu_{ER})^c = {(\nu_1^+ - \nu_2^{-})_L \over \sqrt 2}, \eqno (10)$$
where $\nu_1^+,\ \nu_1^{-}$ are the mass eigenstates.
[Note that the superscripts ($\pm$) refer to the sign under
parity transformation: Under parity, $\nu_1^+ \rightarrow + (\nu_1^+)^c,\
\nu_1^- \rightarrow - (\nu_1^-)^c$.]
Thus, the effect of ordinary matter mixing with
mirror matter is very dramatic. No matter how small
the mass interaction term is, the mixing is maximal.
We have shown this here by the specific case of one
ordinary neutrino and its mirror. This result is
actually more general as we shall see.

This one generation example can easily be extended to three
generations. Under the assumption that intergenerational mixing
is suppressed, which is after all expected considering what
happens in the quark sector,
and is supported by the LSND experiment,
then the three generation case will be, to a first approximation,
3 copies of the mass matrix Eq.(9).
Thus, if nature is described by a parity invariant Lagrangian,
and neutrinos are massive,
then we would expect each of the
three known weak eigenstate neutrinos to be, approximately, maximal
mixtures of two physical states.
Remarkably, the hypothesis that the
electron and the muon neutrinos are
maximal mixtures of two physical states solves
two of the outstanding neutrino puzzles. The solar neutrino deficit
can be explained due to electron neutrino - mirror electron neutrino
oscillation. The atmospheric neutrino anomaly  can be
explained due to muon neutrino - mirror muon neutrino oscillation.
The LSND experiment is also accomodated, due to the
small mixing between the first and second generations.

What are the parameters needed to account for these phenomena?
If we ignore the third generation,
then there will be 4 light states, which will be combinations of the 4 weak
eigenstates:
$\nu_e, \nu_E, \nu_{\mu}, \nu_{M}$ (i.e. the electron neutrino and its mirror,
the
muon neutrino and its mirror).
If there were no intergerational mixing, then these 4 weak eigenstates will
each be
maximally mixed combinations of mass eigenstates of the form:
$$\nu_{eL} = {(\nu_1^+ + \nu_1^{-})_L \over \sqrt 2},$$
$$(\nu_{ER})^c = {(\nu_1^+ - \nu_1^{-})_L \over \sqrt 2},$$
$$\nu_{\mu L} = {(\nu_2^+ + \nu_2^{-})_L \over \sqrt 2},$$
$$(\nu_{
MR})^c = {(\nu_2^+ - \nu_2^{-})_L \over \sqrt 2}.\eqno (11)$$
Of course, in the real world we expect intergenerational mixing,
which means that the weak eigenstates will actually have the form:
$$\nu_{eL} = {\cos\theta \nu_{1L}^+
  \over \sqrt 2} + {\sin\theta \nu_{2L}^+ \over \sqrt 2}
+ {\cos \phi \nu_{1L}^{-} \over \sqrt 2} + {\sin \phi \nu_{2L}^{-}
\over \sqrt 2},$$
$$(\nu_{ER})^c = {\cos\theta \nu_{1L}^+  \over \sqrt 2} +
{\sin\theta \nu_{2L}^+
 \over \sqrt 2}
- {\cos \phi \nu_{1L}^{-} \over \sqrt 2} -
{\sin \phi \nu_{2L}^{-} \over \sqrt 2},$$
$$\nu_{\mu L} = {-\sin\theta \nu_{1L}^+  \over \sqrt 2} +
{\cos\theta \nu_{2L}^+ \over \sqrt 2}
- {\sin \phi \nu_{1L}^{-} \over \sqrt 2} +
{\cos \phi \nu_{2L}^{-} \over \sqrt 2},$$
$$(\nu_{MR})^c = {-\sin\theta \nu_{1L}^+  \over \sqrt 2} +
{\cos\theta \nu_{2L}^+
 \over \sqrt 2}
+ {\sin \phi \nu_{1L}^{-} \over \sqrt 2} -
{\cos \phi \nu_{2L}^{-} \over \sqrt 2},
\eqno (12) $$
where we have assumed that the mass matrix is real.
The assumption that the mass matrix is real
means that $\nu_1^+ $ states can only mix with
$\nu_2^+ $ states, since a mixing term $m\bar \nu_1^+ \nu_2^-$
is forbidden  by parity invariance if $m$ is real. Similarly
the $\nu_1^-$ only mix with the $\nu_2^-$ states.
In general, the weak eigenstates are maximal mixtures
of the mass eigenstates states whether the mass matrix
is real or complex. This can be understood by observing that
the mass eigenstate field must also
be a parity eigenstate otherwise parity would be broken.
The most general form for the mass eigenstate field,
assuming two generations, is:
$$\psi_L = \alpha \nu_{eL}+ \beta (\nu_{ER})^c + \gamma \nu_{\mu L}
+ \delta (\nu_{MR})^c, \eqno (13)$$
where $|\alpha|^2 + |\beta|^2 + |\gamma|^2 + |\delta|^2 = 1$
Requiring that the mass term $m\bar \psi \psi$
[where $\psi \equiv \psi_L + (\psi_L)^c$],
be invariant under parity, requires $\alpha = \beta^*$ and
$\gamma = \delta^*$ which means that
$|\alpha| = |\beta|$ and $|\gamma| = |\delta|$.
This means that in general,
a mass eigenstate has a $1/2$ probability of interacting
like ordinary weak eigenstates $\nu_{e}$ or $\nu_{\mu}$ and
$1/2$ probability of interacting like mirror weak
eigenstates $\nu_E$ or $\nu_M$.
This result can easily be extended to any number of generations.

Thus restricting the mass matrix to be real does not matter as
far as maximal mixing is concerned. This point was not
fully understood in Ref.[15]. In Ref.[15] it was shown that
under the assumption that the mass matrix was real, there is maximal
mixing of the ordinary and mirror weak eigenstates with respect to the
mass eigenstates. However, we have shown here that
the reality condition is unnecessary. Maximal mixing
of ordinary and mirror matter is completely generic. It is an
automatic consequence of the unbroken parity symmetry.

In the particular case of a mass matrix involving only the
minimal particle content of three ordinary left-handed fields and
their mirror partners, the only phases which cannot be absorbed
into the fields can be moved onto the intergenerational mass mixing
terms, and thus, they do not have an important impact on the
physics when the intergenerational mixing is
small [25]. The assumption of a real mass matrix, has the advantage
of simplicity, in that only two parameters $\theta, \phi$ are
required to parameterize the intergenerational mixing (in the simple
two generation case).

If the four mass eigenstates ($\nu_1^+, \nu_1^{-}, \nu_2^+, \nu_2^-$)
have masses
$m_1^+, m_1^{-}, m_2^+, m_2^{-}$, then the
parameters required to explain the solar neutrino deficit,
atmospheric neutrino anomaly and LSND  experiment
are:
$$ 3 \times 10^{-10}\ eV^2 \stackrel{<}{\sim}
|\Delta m_1^2 | \equiv
|m_1^{+2}- m_1^{-2}| \stackrel{<}{\sim}\ 10^{-3 }\ eV^2$$
$$|\Delta m_2^2 | \equiv |m_2^{+2} - m_2^{-2}|\simeq 10^{-2}\ eV^2$$
$$10^{-1}\ eV^2 \stackrel{<}{\sim} |m_2^{+2} - m_1^{+2}|  \stackrel{<}{\sim}
\ 3\ eV^2$$
and
$$ \left(\sin 2\theta + \sin 2\phi\right)^2 \sim 3 \times 10^{-2} -
10^{-3} \eqno (14) $$
respectively.

The range for $|m_1^{+2} - m_1^{-2}|$ is obtained by noting
that we must average
the oscillation probability by taking into account the region of emission
of the sun, the region of absorption on Earth, and
the energy spectrum of the source.
In particular, for the case of 2 state maximal mixing,
the averaged oscillation probablity is 1/2, and is applicable
for $|\Delta m_1^2 | \stackrel{>}{\sim}
3 \times 10^{-10} eV^2$ [26]. The bound
$|\Delta m_1^2 | \stackrel{<}{\sim}
10^{-3} eV^2$ comes from the atmospheric
neutrino anomaly because we must require that the number of electron
neutrinos should not be depleted in that experiment.  Note that there
is a Laboratory bound of
$|\Delta m_1^2 | \stackrel{<}{\sim} 10^{-2}$ eV$^2$[27].

For the atmospheric neutrino anomaly, a recent
analysis of higher energy events
has found a flux dependence on the azimuthal angle [11].
This allows a determination
of the mass difference which is
$|\Delta m^2_2 | \simeq 10^{-2} eV^2$
and the mixing has also been measured to be maximal,
or nearly so ($\sin2\psi   \stackrel{>}{\sim} 0.7$).
The range of parameters for $\Delta m_1^2$ and $\Delta m_2^2$
to explain the solar neutrino deficit and atmospheric neutrino
anomaly, together imply that $|m_2^{+2} - m_1^{+2}| \simeq
|m_2^{-2} - m_1^{-2}|$. This has been used in Eq.(14) to express
the range of parameters suggested by LSND as a constraint on
$|m_2^{+2} - m_1^{+2}|$ only.

Note that in order to explain the atmospheric neutrino anomaly
and the LSND result, it is
necessary for $|\Delta m_2^2| \ll m_2^{\pm 2}$ [28].
This is actually not unexpected since this hierarchy is achieved
if the mass mixing term $m'\bar \nu_{eL} \nu_{ER}$ is much less than
the diagonal terms $m\bar \nu_{eL}(\nu_{eL})^c + (e \leftrightarrow E)$.
This is not unexpected, since it is one
way to understand the CKM matrix in the quark sector. The CKM matrix
is approximately diagonal, which can have a natural origin in
non-diagonal masses being suppressed relative to the diagonal
ones. If this also happens in the lepton sector, not just
between different generations, but also between ordinary and
mirror neutrinos, then it is natural to expect $m'\ll m$, and hence
$|\Delta m_2^2| \ll m_2^{\pm 2}$.

We emphasise that if intergenerational mixing of the
neutrinos is suppressed, then
the exact parity symmetric model is predictive.
It predicts that the electron neutrino oscillates into
its effectively sterile mirror partner in a maximal way, so that the flux
of solar electron neutrinos will be predicted to be 0.5 that
of the minimal standard model (for a large range of $\Delta m_1^2$).
The exact parity model also predicts that the muon neutrino will
oscillate into its effectively sterile mirror partner
also in a maximal way. In other words $\sin^2 2\psi = 1$
is another prediction.
Both of these predictions are supported experimentally, and
will be scrutinized more closely in the near future as
more data is taken, and more experiments are done. This is
especially true for the atmospheric neutrino anomaly.
Experiments will also be able to determine if the muon
neutrino oscillates into the tau neutrino or a sterile
neutrino (for the parameters of interest for the atmospheric neutrino
anomaly). If it is the tau neutrino, then the explanation
given by the exact parity model is ruled out. Also, if the mixing
is not approximately maximal, the parity model explanation will
also be ruled out. It is also very important to realise that
the planned SNO and Super-Kamiokande experiments will be able
to test whether solar electron neutrinos oscillate into active
or sterile species, given that they can detect neutral current
processes. Our model of course predicts that these experiments
should see a factor of two reduction in both charged {\it and}
neutral current events.

What can we say about the tau neutrino?
As in the case of the electron and muon neutrinos,
the tau neutrino should also be approximately a maximal mixture of
two states. Thus, the exact parity model predicts that the
tau neutrino should also oscillate into the effectively
sterile mirror neutrino, also in a maximal way.
Exact parity symmetry does not impose any restriction on
the squared mass difference, so that the oscillation length
is not theoretically constrained.
However, if the neutrinos follow a hierarchical mass pattern,
as the other fermions do, then the tau neutrino will be
the heaviest neutrino, and the squared mass difference
of the $\nu_3^+, \nu_3^-$ parity eigenstates will probably
be at least as large as the $\Delta m_2^2$ for the second generation
(which is $10^{-2} eV^2$ according to the atmospheric neutrino
anomaly). If this is the case,
then it should be possible to experimentally observe the tau
neutrino oscillate into its mirror partner.
We would also expect small intergenerational mixing between
the tau neutrino and the muon and electron neutrinos.
This will also be possible to test experimentally. In fact, several
existing experiments are currently searching for $\nu_{\tau} -
\nu_{\mu}$ oscillations.
Our model does not give any indication for the tau neutrino mass.
Cosmological arguments suggest that the tau neutrino mass should be
less than about $30$ eV.  This bound comes from demanding that
the relic density of tau neutrinos not violate the energy
density bound of the universe (note that there may also be an
allowed window above about 1 MeV for the tau neutrino mass for which
the tau neutrino decays rapidly enough to be within the cosmological
bound). If charged leptons are anything to go by,
then one might expect $m_{\nu_{\tau}}/m_{\nu_{\mu}} \sim m_{\tau}/m_{\mu}$,
which gives a tau neutrino mass of about 15 eV for a muon neutrino mass
of 1 eV. Obviously such a value would put the tau neutrino in
the range for a hot dark matter candidate.
However, without a predictive scheme for fermion masses it is not
possible to draw any firm conclusions.
\vskip 0.7cm
\noindent
{\large \bf IV A naturalness argument}
\vskip 0.5cm
\noindent
Given that $\Delta m_2^2 \approx 10^{-2} eV^2$, and there is
intergenerational mixing between the first and second generations as
measured by the LSND experiment, then it is possible to calculate,
under some simple assumptions, the contribution
of $\Delta m_1^2$ induced from $\Delta m_2^2$ and the intergenerational mixing.
We assume that the second generation neutrinos are much heavier than
the first generation neutrinos, and the effects of the third
generation can be neglected, at least approximately.
In the $\nu^{\pm}$ basis, the mass matrix has the following
simple form if the mass matrix is real:
$${\cal M} = \left(\begin{array}{cccc}
m_1^{'+}&\delta_1&0&0 \\
\delta_1&m_2^+&0&0\\
0&0&m_1^{'-}&\delta_2\\
0&0&\delta_2&m_2^-
\end{array}\right). \eqno (15)$$
Diagonalising this mass matrix, assuming
the second generation neutrinos are heavier than the first
generation neutrinos, i.e.
$m^{\pm}_1 \ll m^{\pm}_2$ we find
$$m_1^+ = m_1^{'+} - \delta_1^2/m_2^+ ;
\ m_1^-= m_1^{'-} - \delta_2^2/m_2^-.
\eqno (16)$$
Note that there are essentially two contributions to $\Delta m_1^2$.
A contribution which depends on the $\delta_{1,2}$ parameters (and is also
independent of $m_1^{'\pm}$) and there are also terms
which depend on $m_1^{'\pm}$. The term which is independent
of $m_1^{'\pm}$ is calculable with the parameters
identified in Eq.(14). The other terms which depend on
$m_1^{'\pm}$ are unknown, however it would be unnatural
for there to be a
significant cancellation between the contribution
we can calculate and the contribution we cannot
(unless there is some crazy symmetry).
Thus, assuming there is no fine tuning between the
$m_1^{'\pm}$ and $\delta^2/m_2$ contributions,
then $$\Delta m_1^2 \stackrel{>}{\sim}
 \left( {\delta_1^2 \over m_2^+}\right)^2
- \left({\delta_2^2 \over m_2^-}\right)^2\   \   $$
$$\approx \sin^4 \theta \ (\Delta m_2^2) + (\sin^4 \theta -
\sin^4\phi)\  (m_2^-)^2 ,\eqno (17)$$
where $\sin\theta = \delta_1/m_2^+, \ \sin\phi = \delta_2/m_2^-$
(and we have assumed that $m_1^{\pm} \ll m_2^{\pm}$).
The mixing angles $\sin\theta, \sin\phi$ parameterize the
intergenerational mixing between the first and second generation,
and are identical to the $\sin\theta, \sin\phi$ defined in Eq.(12).
Thus, we expect
$$\Delta m_1^2 \stackrel{>}{\sim}   \sin^4\phi \ (\Delta m_2^2),\eqno (18)$$
with $\theta = \phi$. From the LSND experiment, $\sin^2 2\phi > 10^{-3}$,
so that we expect
$$\Delta m_1^2 \stackrel{>}{\sim} 10^{-9}\ eV^2.\eqno (19)$$
Thus, it is interesting that the range of parameters necessary
to solve the atmospheric neutrino anomaly, together with the
intergenerational mixing as measured by the LSND experiment, imply that
the range of parameters for $\Delta m_1^2$ is expected to
be in the range necessary to reduce the flux of solar electron neutrinos
by a factor of 2 (this occurs for $\Delta m_1^2 \stackrel{>}{\sim}
10^{-10} eV^2$).
If there were no solar neutrino deficit, then the exact
parity model would not be able to explain the atmospheric neutrino anomaly
and simultaneously account for the LSND experiment in a compelling way.
A deficit of solar neutrinos appears to be a necessary consequence.

\vskip 0.7cm
\noindent
{\large \bf V The see saw model.}
\vskip 0.5cm
\noindent
Hitherto, we have discussed neutrino oscillations in the parity symmetric model
without focussing on any particular model. This is possible, because parity
symmetry allows us to make the prediction that the weak eigenstates will
be maximal mixtures of two states, independently of the details of where the
masses
come from. We now focus briefly on a particular model.
As is well known, in the standard model neutrinos must be
massless. Thus, if neutrinos have non-zero masses the standard
model must be modified. There are very few good ideas for understanding
the smallness of the neutrino masses relative to the
masses of the other fermions.
The simplest possibility known at the moment is the see-saw
model [29]. This involves assuming the existence of a gauge singlet
right-handed neutrino which develops a large Majorana mass.
The see-saw model is a simple way to understand the
smallness of the masses of the known (i.e. the three left-handed)
neutrinos.
In the usual see-saw model there are
two Weyl neutrino fields per generation. Denote these
by $\nu_L$ and $\nu_R$. The $\nu_L$ field is a member
of a $SU(2)_L$ doublet while $\nu_R$ is a gauge singlet.
The usual Higgs doublet can couple the $\nu_L$ and $\nu_R$
together and its vacuum expectation value will generate a
Dirac mass term. Also, since we assume that $\nu_R$ is electrically
neutral it can have a bare Majorana mass term coupling it
to itself.
Thus we have two mass terms:
$${\cal L}_{mass} = 2m\bar \nu_L \nu_R +
M \bar \nu_R (\nu_R)^c + H.c. \eqno (20)$$
Note that $M$ is a bare mass term, and can take any value, while
$m$ is a mass term which is generated when the electroweak
gauge symmetry is broken. It is usually assumed that $M \gg m$,
since $M$ is not protected by the gauge symmetry.
The mass matrix has the form:
$${\cal L}_{mass} = [\bar \nu_L,  (\bar \nu_R)^c]
\left(\begin{array}{cc}
0&m\\
m&M
\end{array}\right)\left[
\begin{array}{c}
(\nu_L)^c\\
\nu_R
\end{array}\right] + H.c. \eqno (21)$$
Diagonalising this mass matrix yields two Majorana mass eigenstates
with masses $m^2/M$ and $M$ (assuming that $M \gg m$).
If we denote the mass eigenstates by $\nu_{light}$ and $\nu_{heavy}$
then they can be written in terms of the weak eigenstates as
follows:
$$\nu_{light L}= \cos\phi \ \nu_L + \sin\phi \ (\nu_R)^c,$$
$$\nu_{heavy R} = - \sin\phi \ (\nu_L)^c + \cos\phi \ \nu_R,
\eqno (22)$$
where $\tan\phi = m/M$. Thus in the limit $M \gg m$,
we see that the light state is
essentially $\nu_L$ while the heavy state is essentially $\nu_R$.

The see-saw model is a simple extension of the standard model.
As in the case of
the standard model, it is straightforward to make it exactly
parity invariant [15]. In this case, there are
four Weyl neutrino fields per generation:
$\nu_{L}, \nu_{R}$ and their mirror images $ N_{R}, N_{L}$.
Under the parity symmetry $\nu_{L,R} \leftrightarrow \gamma_0
N_{R,L}$. Note that since $N_R$ is the parity partner of $\nu_L$
it belongs to an $SU(2)_2$ doublet; while $N_L$ being the parity
partner of $\nu_R$ is a gauge singlet. [Recall that
the gauge group is defined in Eq.(3) and the parity transformations
are given in Eq.(5).] Assuming the minimal Higgs sector of one
ordinary Higgs doublet and its mirror image,
then the following mass terms are allowed (where for simplicity
we examine only one generation):
$${\cal L}_{mass} = 2m_1(\bar \nu_L \nu_R + \bar N_R N_L) +
2m_2 [\bar \nu_L (N_L)^c + \bar N_R (\nu_R)^c]$$
$$+ M_1 [\bar \nu_R (\nu_R)^c + \bar N_L (N_L)^c]
+ M_2 (\bar \nu_R N_L + \bar N_L \nu_R) + H.c.
\eqno (23)$$
Note that $m_{1,2}$ are mass terms which arise from
spontaneous symmetry breaking, while $M_{1,2}$ are bare mass terms.
As in the case discussed above, we will assume that $M_{1,2} \gg
m_{1,2}$. We will first also assume that the masses are real.
This is not an important restriction and it is
the only way in which we depart from the most general case.
Later we will coment on the general complex case. From Eq.(23) we
see that the mass matrix has the form:
$${\cal L}_{mass} = \bar \nu_L {\cal M} (\nu_L)^c + H.c.,
\eqno (24)$$
where
$$\nu_L = [(\nu_L)^c,  N_R,  \nu_R, (N_L)^c]^T, \eqno (25)$$
and
$${\cal M} = \left(\begin{array}{cccc}
0&0&m_1&m_2\\
0&0&m_2&m_1\\
m_1&m_2&M_1&M_2\\
m_2&m_1&M_2&M_1
\end{array}\right). \eqno (26)$$
The mass matrix can be simplified by changing to the parity
diagonal basis $\nu^{\pm}_{L} = {\nu_L \pm (N_R)^c \over
\sqrt{2}}$, and $\nu^{\pm}_{R} = {\nu_R \pm (N_L)^c \over
\sqrt{2}}$.
In this basis the mass matrix has the form
$${\cal M} = \left(\begin{array}{cccc}
0&0&0&m_{+}\\
0&0&m_{-}&0\\
0&m_{-}&M_{-}&0\\
m_{+}&0&0&M_{+}
\end{array}\right), \eqno (27)$$
where $m_{\pm} = m_1 \pm m_2 $ and
$M_{\pm} = M_1 \pm M_2 $ .
The mass matrix can now be easily diagonalised because
it is essentially two copies of the $2 \times 2$
mass matrix Eq.(21). In the limit $M_{\pm} \gg m_{\pm},$ the
mass matrix Eq.(27) has eigenvalues:
$$m_{+}^2/M_{+}, m_{-}^2/M_{-}, M_{+}, M_{-},
\eqno (28)$$
and eigenvectors (which are the mass eigenstates):
$$\nu_1 = \nu^{+}_{L}, \  \nu_2 = \nu^{-}_L,\
\nu_3 = \nu^{+}_R, \   \nu_4 = \nu^{-}_R. \eqno (29)$$
If we had started with Eq.(26) being the most general complex
matrix, then Eq.(29) would still follow.
To see this, observe that the heavy states with masses $M_{\pm}$, are
approximately mixtures of the $\nu_R$ and $N_L$ fields, with
small admixture of $\nu_L$ and $N_R$ fields. Hence,
the light states must be approximately comprised of $\nu_L$
and $N_R$ fields. Since, by parity invariance, mass eigenstates
must also be parity eigenstates (except in the case where the
states are degenerate), it follows that the light states
($\nu_1, \nu_2$) must be approximately of the form given in Eq.(29)
(since only these combinations are parity eigenstates).
Another way of saying this is that the heavy $\nu_R$ and $N_L$
fields decouple, leaving the two light neutrino states
$\nu_L$ and $N_R$, which must have a mass matrix of the form Eq.(9)
by parity invariance. The discussion following Eq.(9) consequently
holds for the see-saw model as well as the minimal model without
gauge singlet neutrinos.

Thus we conclude that in the one generation case, there is effectively only
two state mixing:
$$\nu_L = {\nu^{+}_L + \nu^{-}_L \over \sqrt{2}} $$
$$\  \     = {\nu_1 + \nu_2 \over \sqrt{2}}. \eqno (30)$$
Thus in this case the neutrino oscillation probability
averaged over many oscillations
is $1/2$. In the physical case of three
generations, in general $\nu_L^{+}$ and $\nu_L^{-}$ will each
be linear combinations of mass eigenstates. The details
depend on the precise form of the mass matrix.
However, the assumption that intergenerational mixing
is small means that
the one generation result will be a good approximation,
and that each weak eigenstate ($\nu_{eL}$, $\nu_{\mu L}$, $\nu_{\tau L}$
and their mirror partners) will each be approximately a
maximally mixed combination of mass eigenstates.

The see-saw model illustrates the results of the previous sections
in a concrete model. Similar results should also occur in any
other model for neutrino masses, provided that there are nonzero
neutrino-mirror neutrino mass mixing terms.

\vskip 0.7cm
\noindent
{\large \bf VI Exact charge conjugation invariance?}

\vskip 0.5cm
\noindent
One variant of the exact parity model is an exact (unorthodox)
charge conjugation invariant model. This is very similar to the
exact parity model. The only difference is that
the mirror particles are assumed to have the {\it same}
chirality as the ordinary fermions
(recall in the exact parity model, the mirror fermions have
the opposite chirality to the ordinary fermions).
Explicitly,
the gauge symmetry of the theory is
$$SU(3)_1 \otimes SU(2)_1 \otimes U(1)_1 \otimes
SU(3)_2 \otimes SU(2)_2 \otimes U(1)_2. \eqno (31)$$
and the fermions consist of the ordinary particles
and their C images, which transform under the
gauge group of Eq.(31) as
$$\begin{array}{ll}
f_L \sim (1, 2, -1)(1, 1, 0),& F_L \sim (1, 1, 0)(1, 2, -1),\\
e_R \sim (1, 1, -2)(1, 1, 0),& E_R \sim (1, 1, 0)(1, 1, -2),\\
q_L \sim (3, 2, 1/3)(1, 1, 0),&Q_L \sim (1, 1, 0)(3, 2, 1/3),\\
u_R \sim (3, 1, 4/3)(1, 1, 0),&U_R \sim (1, 1, 0)(3, 1, 4/3),\\
d_R \sim (3, 1, -2/3)(1, 1, 0),&D_R \sim (1, 1, 0)(3, 1, -2/3),
\end{array}
\eqno (32)
$$
(with generation index suppressed). The Lagrangian is invariant
under the discrete $Z_2$ unorthodox C symmetry defined by
$$\begin{array}{c}
G_1^{\mu} \leftrightarrow G_2^{\mu},\
W_1^{\mu} \leftrightarrow W_2^{\mu},\ B_1^{\mu}
\leftrightarrow B_2^{\mu}\\
f_L \leftrightarrow F_L, \ e_R \leftrightarrow E_R, \\
q_L \leftrightarrow Q_L, \ u_R \leftrightarrow U_R\
d_R \leftrightarrow D_R
\end{array}
\eqno (33)$$
As in the exact parity model, there are
two Higgs multiplets, $\phi_1$ and
$\phi_2$, which are partners under the discrete symmetry [the Higgs
potential is the same as in the exact parity model, see Eq.(7)].
We denote this $Z_2$ symmetry unorthodox C symmetry, since it
is essentially unorthodox parity times ordinary CP
[after relabeling $(F_R)^c $ as $F_L$ etc.].
If ordinary CP were conserved,
then the exact parity model would also be an exact unorthodox C invariant
model. However ordinary CP is violated, so that the exact
unorthodox C invariant model is similar but not exactly the same
as the exact parity invariant model.
Its testible predictions are also similar to the exact parity
symmetric model, however they are not exactly the same.
In particular, note that in general, even for complex mass
matrices, the unorthodox C eigenstates $\nu^+$ and $\nu^-$
cannot be coupled together with a mass term. This is because
$\bar \nu^+ \nu^- \rightarrow - \bar \nu^+ \nu^-$ under C
transformation.  Recall that in the exact parity symmetric model,
$\nu^+$ and $\nu^-$ could mix together if the mass matrix
was complex since under parity, $\bar \nu^+ \nu^- \rightarrow
\bar{(\nu^+)^c } (\nu^-)^c$. The two cases should be physically
distinct if the mass matrix is complex, and this issue
should be studied in order to determine precisely how
the exact parity and exact C symmetric models could be
differentiated experimentally. This issue however, we leave
for future work.
\vskip 0.7cm
\noindent
{\large \bf VII Conflict with standard big bang nucleosynthesis?}
\vskip .5cm
\noindent

The standard scenario of Big Bang Nucleosynthesis (BBN) can put
constraints on the energy density of the universe when it has
temperatures of the order of an MeV and below.
This in turn can bound the
number of relativistic degrees of freedom, which in the standard
scenario comprise photons and neutrinos. Over the last few years,
the upper bound on the number of neutrino flavours $N_{\nu}$ has
steadily decreased as increasingly more accurate astronomical and
nuclear data have become available. Until very recently, these
data were consistent with the SM prediction of $N_{\nu}=3$.
However, a recent analysis [30] argues that the best fit is
obtained with $N_{\nu}=2$, and furthermore that $N_{\nu}=3$
is ruled out at $99.7\%$ C.L. There thus appears to be
an incompatibility between the minimal SM of particle physics
and the standard Hot Big Bang model of cosmology. Because of the
previous apparent success of BBN, it has become standard
practice to use compatibility with BBN to put constraints
on extensions of the SM of particle physics. The new doubts
about BBN also cast doubt on the veracity of this class of
bound on new particle physics. It is possible, for instance,
that Big Bang cosmology rather than the SM of
particle physics will need to be altered because of
this conflict [31].

If the mirror matter exists, then it will also be hard to explain
the primordial abundances of light elements observed in the
universe. This is because one expects three extra neutrino
species, as well as the mirror photon to contribute to the energy
density of the early universe during the nucleosynthesis era. According to the
standard theory, this will cause the universe to expand too
rapidly and leads to unacceptable predictions for light element
abundances (given the standard assumptions). As discussed above, this
is also the case for the SM, but the problem is even more severe
in the exact parity model. Just as it is inappropriate to rule out
the minimal SM from its incompatibility with BBN, it is also
premature to rule out other particle physics models that
do not accord with BBN.

Nevertheless, it is interesting to speculate about how BBN and
the exact parity model might be reconciled.
One might think that this problem could be alleviated if the
temperature of the mirror matter is assumed to be much less than
the temperature of the
ordinary matter. This could be due to some new physics at
very high temperatures [32] or, possibly, divine intervention.
However, the oscillations of the muon neutrino into the mirror
muon neutrino, necessary to solve the atmospheric neutrino anomaly,
would put the mirror neutrino in equilibrium with the ordinary
matter.
This result should follow from the analysis of a sterile neutrino
mixing with the muon neutrino. The bound [33]
$$ \sin^2 2\theta\ \delta m^2 \stackrel{<}{\sim}
10 ^{-6}\ eV^2,\eqno(34)$$
which should be obeyed to prevent sterile species from coming into
thermal equilibrium due to oscillation, is violated by the
parameters necessary to solve the atmospheric neutrino anomaly.
The mirror weak interactions should put the mirror muon neutrino
into equilibrium with the entire mirror sector.
It seems then that either (a) some modification of the
usual nucleosynthesis scenario is required, or (b) some
assumptions that underlie the derivation of Eq.(34) need to
be examined. Some ideas include:
\vskip .3cm
\noindent
1) One important assumption behind Eq.(34) is that the
neutrino-antineutrino asymmetry is
not much larger than the baryon-antibaryon
asymmetry. Suppose instead that the initial $\nu-\overline{\nu}$
asymmetry is much larger than the baryon-antibaryon asymmetry (but
still small enough to negligibly affect the expansion rate of the
universe), for example
$\Delta L \equiv \Delta n_{\nu}/n_{\gamma} \sim 10^{-5}$.
Previous work [33] has shown that for $\Delta L$ below a critical value,
the dynamical evolution of neutrino and antineutrino number densities
reduces $\Delta L$ to zero. If on the other hand $\Delta L$ is above
this critical value, then the initial $\Delta L$ can persist [34, 35].
A nonzero $\Delta L$ can severely suppress the transition rate from
active to sterile neutrinos. For such large values of $\Delta L$
the bound Eq.(34) does not apply, and reconciliation between the
sterile neutrino solution to the atmospheric neutrino anomaly
is possible [35].
\vskip .3cm
\noindent
2) There could exist a large {\it negative} cosmological constant at
the nucleosynthesis era, which would slow down the expansion
rate of the universe at early times, which would have the
opposite effect to increasing the number of neutrino species.
We do not know if such a possibility has been seriously considered
in the literature before, but it seems interesting to us. Of course,
the cosmological constant today would have to be much smaller
than at the time of nucleosynthesis. Such a time-dependent
cosmological constant may be difficult to implement in a natural
manner.
\vskip .3cm
\noindent
3) The non-zero neutrino masses could be due to some
new electroweak symmetry breaking mechanism
at some quite low scale ($\stackrel{<}{\sim}$ 1 MeV). This
symmetry breaking scale
could be associated with a phase transition in the early universe,
so that the neutrinos are effectively massless at temperatures
above about 3 MeV (i.e. the temperature where the neutrinos go out
of thermal equilibrium).
If this happens, then in the early universe the oscillations
do not occur and the mirror sector does
not come into equilibrium with the ordinary sector (at least
not through the mechanism of neutrino oscillations). Thus, a
temperature difference between the ordinary and mirror worlds
could be maintained if it was set up doing very early times
due to physics at extremely high energies.
\vskip .3cm
\noindent
4) It has been shown that if the tau neutrino is very heavy,
between 1 and 10 MeV, then the total energy density in the early universe
can be doubled without conflict with nucleosynthesis
provided that the tau neutrino decays in a particular range of lifetimes
and includes electron neutrinos in its decay products [36, 37, 38].
The electron neutrinos in the decay of tau neutrinos
convert neutrons into protons, which has the opposite effect
to increasing the energy density in the early universe.
\vskip .3cm
\noindent
5) It is rather theoretically appealing to have parity symmetry
unbroken by the vacuum. However it is worthwhile, in view
of the nucleosynthesis difficulties, to mention another
possibility.
It is possible that the parity symmetry in the exact parity
symmetric model is slightly broken [39]. In this case, the large
angle neutrino oscillations would no longer be an automatic
consequence of the parity symmetry. However, if the symmetry
is only broken slightly then large angle neutrino oscillations
would still occur for a large range of parameters.
Below we illustrate how having parity slightly broken can
lead to a model with acceptable
nucleosynthesis predictions.

If we assume
that the parity symmetry is spontaneously broken in
such a way that the mirror photon is massive (i.e.
mirror electromagnetism is spontaneously broken) [40],
and heavier then the mass of the mirror electron and
mirror positron, then this means that the mirror photon will be unstable
and rapidly decay into mirror electron-positron pairs.
If we also assume that the mirror electron
is still the lightest charged
mirror fermion, but its mass is changed so
that it is slightly heavier that the ordinary
electron ($ m_e < m_E \stackrel{<}{\sim} 3$ MeV), then
the mirror electron-positron pair can annihilate into
ordinary electrons and positrons via an intermediate
mirror photon [the ordinary electron-positron pair can
interact with the mirror photon because of the $U(1)$ kinetic
mixing term] [41].
This will be the dominant annihilation channel since we are
assuming that the annihilation into mirror photons is not
kinematically allowed.

Thus, in the early universe, when the temperature drops to a few
MeV, the mirror electrons and positrons will begin to annihilate
into the ordinary electrons and positrons. Thus, if they were
in equilibrium before this time (which is what we would expect),
then the annihilation of the mirror electrons and mirror positrons
will heat the
ordinary electrons, positrons and ordinary photons (but
not the neutrinos, since they have already decoupled).
These particles will be hotter than the neutrinos because of
this heating. Below the temperature of 1 MeV, the ordinary
electrons and positrons have disappeared as well, leaving
only the photons and neutrinos (3 ordinary neutrinos as well
as 3 mirror neutrinos). Using the usual methodology (i.e.
conservation of entropy),
we can calculate the temperature of the neutrinos relative
to the photons, which is
$$T_{\nu} = \left({2 \over 9}\right)^{1\over 3}T_{\gamma}.
\eqno (35)$$
Note that the temperature of the neutrinos is significantly
less than the usual, standard value of
$(4/11)^{1 \over 3} T_{\gamma}$
because of the mirror electron - mirror positron annihilation.
Thus, the energy density of the neutrinos is reduced because
of this reheating by a factor
$$\left( { 2 \over 9}{11 \over 4}\right)^{4 \over 3} =
\left({11 \over 18}\right)^{4 \over 3} \simeq 0.52 \eqno (36)$$
where we have used the fact that the energy density is proportional
to the fourth power of the temperature. However, this reduction in
energy density is compensated because
there are twice as many neutrino species (3 ordinary plus 3
mirror neutrinos).
Thus, we end up with an effective energy density of neutrinos which
is essentially the same as the standard model case
(since $2 \times 0.52 \simeq 1$).
\vskip .3cm
\noindent
6) Finally, note that the suggestions (1)-(5) all involve working
within the standard big bang model of cosmology.
It could be that a more radical modification of cosmology
is required. One should keep in mind that the standard cosmological
model, although simple, and quite successful, also has many
open problems. Due to the nature of cosmology, it is difficult
to rigorously test the standard theory and it is possible
some fundamental changes could be necessary. We should keep
an open mind [42].
\vskip .2cm

While the exact parity model has problems explaining the light
element abundances within the framework of the standard big bang
model, there are other observations
which can be viewed as evidence in favour of the exact parity
model. There is evidence on both large and small scales
that there is additional matter, called dark matter which has
so far escaped direct observation. The exact parity model
provides a candidate for the dark matter, which is matter
comprised of mirror particles [19].

Finally, note that one cosmological advantage of having parity
unbroken as opposed to spontaneously broken is that there will
be no domain walls. Domain walls tend to be a problem for
theories with spontaneously broken symmetries [43], since the
energy density of domain walls can overclose the universe.

\vskip 0.7cm
\noindent
{\large \bf VIII Conclusions}
\vskip 0.5cm
\noindent
The exact parity symmetry is a theoretically
appealing symmetry beyond the standard model. It is also
experimentally appealing, since it predicts the existence
of additional light neutrino species, which, due to the
parity symmetry, automatically lead to a large angle neutrino
oscillations. These oscillations are consistent with
the observations of atmospheric neutrinos and the solar neutrino
flux.

Thus, the main conclusion is that the atmospheric neutrino anomaly,
the solar neutrino deficit and the LSND experiment are all
consistent with the predictions/expectations of the
standard model extended to include an exact unbroken parity
symmetry. This explanation will be tested more rigorously in the near future
as more data is analysed from existing and several new experiments
(SNO and Super-Kamiokande for instance).
It is a remarkable prospect that exact parity invariance may
have to be reconsidered as a serious candidate for an exact
symmetry of nature. We eagerly await the experiments!

\vskip 1cm
\noindent
{\large \bf Note Added}

After placing this manuscript on the e-print archive hep-ph, a
paper by Z. Berezhiani and R. N. Mohapatra appeared (hep-ph/9505385)
which also
deals with a mirror matter model in the context of neutrino
anomalies, but in a different way.
\vskip 1cm
\noindent
{\large \bf Acknowledgements}
\vskip .5cm
\noindent
This work was supported by the Australian Research Council.

\vskip 1.5cm
\noindent
{\large \bf References}
\vskip .7cm
\noindent
[1] LSND Collaboration, Preprint, nucl-ex/9504002.
\vskip .5cm
\noindent
[2] R. Foot, Mod. Phys. Lett. A6, 527 (1991). For a review of Electric charge
quantization in the standard model, see R. Foot, G. C. Joshi, H. Lew,
and R. R. Volkas, Mod. Phys. Lett. A5, 2721 (1990) and J. Phys.
G{\bf 19}, 361 (1993).
\vskip .5cm
\noindent
[3] R. Davis, in Proceedings of the 21st International cosmic
ray conference, ed. R. J. Protheroe (University of
Adelaide Press, Australia, 1990) Vol.2 p 143;
Kamiokande collabroration, Y. Suzuki, Talk at the TAUP 93 Conference,
Gran Sasso, Sept. 1993.
GALLEX Collaboration, P. Auselmann et al; Phys. Lett.
B327, 377 (1994); SAGE Collaboration,
J. Abdurashitov et. al. Phys. Lett. 328B, 234 (1994).
(1991).
\vskip .5cm
\noindent
[4] J. N. Bahcall and M. H. Pinsonneault, Rev. Mod. Phys. 64, 885 (1992).
\vskip .5cm
\noindent
[5] S. Turck-Chieze and I. Lopes, Astrophys. J. 408, 347 (1993).
\vskip .5cm
\noindent
[6] D. R. O. Morrison, Int. J. Mod. Phys. D1, 281 (1992);
CERN preprint/93-98 (1993).
\vskip .5cm
\noindent
[7] B. Pontecorvo, Sov. JETP 26, 984 (1968);
V. Gribov and B. Pontecorvo, Phys. Lett. B28, 493 (1969).
\vskip .5cm
\noindent
[8] Our point of view is not universally accepted, and some
authors appear to disagree with our point of view [see for example,
P. I. Krastev and S. T. Petcov, Phys. Rev. Lett. 72, 1960 (1994)],
and argue that the discrepancy between the
chlorine experiment and the other three experiments is
evidence that there is a significant energy dependence to the observed
discrepancy.
We feel that this is not clear at the
moment.Note that the discrepancy depends a lot
on which standard solar model you use. For the solar
model of Turck-Chieze et al.,
the ratio of Homestake to Kamiokande data (normalised to the prediction
of Turck-Chieze et al.), taken at the same time
(i.e. since 1988) is $R_H/R_K = 0.70 \pm 0.09 \pm 0.07$
without theoretical uncertainties (see S. Turck-Chieze et. al.,
Phys. Rep. 230, 57 (1993)).
This ratio is consistent
with 1. In other words, at the present time, there is no
statistically compelling evidence that there is a
{\it energy dependent} flux deficiency.
Even if there is a discrepancy, it may
be due to many sources,
e.g. the absorption cross section for $^{37}Cl$
is not well known etc.
\vskip .5cm
\noindent
[9] K. Hirata et al, Phys. Lett. B280,
146 (1992);
IMB Collaboration, D. Casper et al, Phys. Rev. Lett. 66, 2561 (1991);
R. Becher-Szendy et al., Phys. Rev. D46, 3720 (1992).
\vskip .5cm
\noindent
[10] T. K. Gaisser, T. Stanev and G. Barr, Phys. Rev. D38, 85 (1988);
G. Barr, T. K. Gaisser and T. Stanev, Phys. Rev. D39, 3532 (1989).
\vskip .5cm
\noindent
[11] Kamiokande Collaboration, Y. Fukuda, Phys. Lett. 335B, 237
(1994).
\vskip .5cm
\noindent
[12] Light neutrinos can be members of exotic
$SU(2)_L \otimes U(1)_Y$ multiplets
if they have $I_3 = Y = 0$ and consequently do not couple to the Z-boson.
For a model of this type see e.g.
R. Foot, H. Lew, and G. C. Joshi, Phys. Lett. 212B, 67 (1987).
\vskip .5cm
\noindent
[13] T. D. Lee and C. N. Yang, Phys. Rev. 104, 254  (1956).
\vskip .5cm
\noindent
[14] R. Foot, H. Lew and R. R. Volkas, Mod. Phys. Lett. A7, 2567
(1992).
\vskip .5cm
\noindent
[15] R. Foot, Mod. Phys. Lett. A9, 169 (1994).
\vskip .5cm
\noindent
[16] R. Foot, H. Lew and R. R. Volkas, Phys. Lett. B272, 67 (1991).
\vskip .5cm
\noindent
[17] V. Kobzarev, L. Okun and I. Pomeranchuk, Sov. J. Nucl. Phys.
3, 837 (1966); A. Salam, Nuovo Cimento, 5, 299 (1957);
M. Pavsic, Int. J. Theor. Phys. 9, 229 (1974);
\vskip .5cm
\noindent
[18]  S. L. Glashow, Phys. Lett. B167, 35 (1986); E. D. Carlson
and S. L. Glashow, Phys. Lett. B193, 168 (1987); see also
B. Holdom, Phys. Lett. B166, 196 (1985).
\vskip .5cm
\noindent
[19] E. W. Kolb, D. Seckel and M. S. Turner, Nature 514,
415 (1985); H. M. Hodges, Phys. Rev. D47, 456 (1993).
\vskip .5cm
\noindent
[20] There are many examples of models with a
exact parity symmetric Lagrangian,
which is spontaneously broken by the vacuum.
These include models based on
$SU(4) \otimes SU(2)_L \otimes SU(2)_R$
[J. C. Pati and A. Salam, Phys. Rev.D10, 275 (1974)],
and  $SU(3)_c \otimes SU(2)_L \otimes SU(2)_R \otimes U(1)_{B-L}$
[R. N. Mohapatra and
J. Pati, Phys. Rev. D11, 566, 2558 (1975);
G. Senjanovic and R. N. Mohapatra, Phys. Rev. D12, 1502 (1975);
R. Mohapatra and G. Senjanovic, Phys. Rev. D23, 165 (1981)].
These models have parity symmetry interchanging the
$SU(2)_L $ gauge bosons with
$SU(2)_R$ (along with $x \leftrightarrow -x$ of course).
There is no fundamental
reason why parity should interchange $SU(2)_L$
with $SU(2)_R$, and there are other
parity symmetric models which do not exhibit this feature.
These include models
based on $SU(3)_c \otimes SU(3)_l \otimes SU(2)_L \otimes U(1)$
[R. Foot and H. Lew,
Phys. Rev. D41, 3502 (1990); R. Foot, H. Lew and R. R. Volkas,
Mod. Phys. Lett. A8, 1859 (1992)]
in which parity symmetry interchanges the
$SU(3)_c$ of QCD with a $SU(3)$ of leptonic
colour (quark-lepton symmetric models).
Another unorthodox parity symmetric model has the
parity operation interchanging
the $SU(3)$ of QCD with a  $SU(3)_L$ which
contains the usual weak interactions i.e.
the gauge group of the model is
$SU(3)_c \otimes SU(3)_L \otimes U(1)$ [R. Foot,
Mod. Phys. Lett. A10, 159  (1995).]
There are also spontaneously broken parity
symmetric models which have a mirror universe.
For a model of this form with gauge group,
$SU(3) \otimes SU(2) \otimes SU(2)
\otimes U(1) \otimes U(1)$
see S. M. Barr, D. Chang, and G. Senjanovic, Phys. Rev. Lett. 67, 2765 (1991).
For another model of this type, see R. Foot and H. Lew in Ref. [21].
While there are many different types of spontaneously
broken parity symmetric models,
there is essentially only one type of unbroken
parity symmetric models [modulo trivial
extensions, e.g. $SU(5)_{gut} \otimes SU(5)_{gut}$].
\vskip .5cm
\noindent
[21] In general this Higgs potential can only have
two possible vacua; $\langle \phi_1 \rangle = \langle \phi_2 \rangle,$ or
$\langle \phi_1 \rangle = u, \langle \phi_2 \rangle =0$. This second vacuum
corresponds to the region of parameter space
$\lambda_1 + \lambda_2 > 0, \  \lambda_2 < 0$.
The resulting model has
been discussed by R. Foot and H. Lew, McGill
University, Taiwan institute preprint, hep-ph/9411390 (1994).
\vskip .5cm
\noindent
[22] B. Holdom (in Ref. [18]) pointed out that a kinetic mixing term
could be radiatively induced by assuming the existence of
heavy particles which have both ordinary and miror charges.
Such a mechanism for generating kinetic mixing
is required if the gauge group does not contain $U(1)$ factors
(as is the case in grand unified models, for example).
Holdom observed that kinetic mixing can cause the mirror particles
to couple to the ordinary photon, thus giving them effectively
a mini electric charge. Independently of Holdom's work, we also
observed this feature of kinetic mixing in the context of a
particular model, the parity conserving model (see Ref.[16]).
Following R. Foot and X.-G. He, Phys. Lett.
B267, 509 (1991) [which seems to be the first place where kinetic mixing
terms were proposed as a possible {\it tree level} term in a gauge
theory with two $U(1)$ factors],
we observed that the kinetic mixing term can occur at tree level,
and should be viewed as a, a priori, fundamental parameter of
the theory.
There is no need to assume the existence of additional heavy particles.
\vskip .5cm
\noindent
[23] E. Golowich and R. W. Robinett, Phys. Rev. D35, 391 (1987);
S. Davidson, B. Campbell, and D. Bailey, Phys. Rev. D43, 2314
(1991); M. I. Dobroliubov and A. Yu. Ignatiev, Phys. Rev. Lett. 6, 679
(1990).
\vskip .5cm
\noindent
[24] Science, 267, 1424 (1995).
\vskip .5cm
\noindent
[25] We may also need to assume that there is no degeneracy
among the neutrinos belonging to different generations.
\vskip .5cm
\noindent
[26] J. N. Bahcall, {\it Neutrino astrophysics} (Cambridge University
Press, Cambridge, 1989).
\vskip .5cm
\noindent
[27] Particle data group, Phys. Rev. D{\bf 50}, 1173 (1994).
\vskip .5cm
\noindent
[28] Assuming that $m_2^{\pm} > m_1^{\pm}$. If $m_1^{\pm} > m_2^{\pm}$
then Eq.(14) implies that $|\Delta m_1^2 | \ll m_1^{\pm 2}$.
Thus, either $|\Delta m_2^2| \ll m_2^{\pm 2}$ or $|\Delta m_1^2|
\ll m_1^{\pm 2}$ is required.
\vskip .5cm
\noindent
[29] M. Gell-Mann, P. Ramond, and R. Slansky, in Supergravity,
Proceedings
of the Workshop, Stony Brook, New York, 1979, edited by
P. van Nieuwenhuizen and D. Z. Freedman (North Holland,
Amsterdam, 1979);
T. Yanagida, Proceedings of the workshop on unified
theories and baryon number in the universe, Tsukuba,
Japan, 1979, edited by A. Sawada and A. Sugamoto (KEK
Report No. 79-18, Tsukuba-Gun, Ibaraki-Ken, Japan, 1979);
R. N. Mohapatra and G. Senjanovic, Phys. Rev. Lett. 44, 912 (1980).
\vskip .5cm
\noindent
[30] N. Hata et al., Ohio State University report No.\ OSU-TA-6/95,
hep-ph/9505319.
\vskip .5cm
\noindent
[31] It is of course also possible that the SM of particle physics
needs modification. One idea is to extend the theory so that
massive tau-neutrinos decay rapidly.
\vskip .5cm
\noindent
[32] See, E. Kolb et al. in Ref. [19] for one such scenario.
\vskip .5cm
\noindent
[33] A. Dolgov, Sov. J. Nucl. Phys. 33, 700 (1981);
R. Barbieri and A. Dolgov, Nucl. Phys. B349, 743 (1991);
K. Kainulainen, Phys. Lett. 244B, 191 (1990); K. Enqvist,
K. Kainulainen and M. Thomson,
Nucl. Phys. B373, 191 (1992);
J. Cline, Phys. Rev. Lett. 68, 3137 (1992);
X. Shi et. al. Phys. Rev. D48, 2568 (1993).
\vskip .5cm
\noindent
[34] K. Enqvist, K. Kainulainen and J. Maalampi, Phys. Lett. B244, 186
(1990).
\vskip .5cm
\noindent
[35] R. Foot and R. R. Volkas, University of Melbourne report UM-P-95/75,
hep-ph/9508275, Phys. Rev. Lett. (in press).
\vskip .5cm
\noindent
[36] A. D. Dolgov and D. P. Kirilova, Int, J. Mod. Phys. A3, 267 (1988);
N. Terasawa and K. Sato, Phys. Lett. B185, 412 (1988).
\vskip .5cm
\noindent
[37] S. Dodelson, G. Gyuk, M. S. Turner, Phys. Rev. Lett. 72, 3754 (1994).
\vskip .5cm
\noindent
[38] Actually, a heavy tau neutrino could decay quite rapidly
in the exact parity symmetric model, since
there is no GIM mechanism for the neutrinos, and the decay:
$\nu^+_3 \rightarrow \nu_3^- + Z^*$ (where $Z^*$ is a virtual
Z boson (actually it is the parity diagonal combination
$(Z_1 - Z_2)/\sqrt 2 $), and it  will decay
into $\nu_e, \nu_E, e^{\pm}$, etc.) is
not suppressed by any small mixing angle. A rough calculation
suggests that there is a small window of $m_{\nu_{\tau}}$ between
$9$ and $10$ MeV, where the lifetime is within the bound quoted
in Ref [37].
\vskip .5cm
\noindent
[39] To build a model with parity symmetry slightly broken in the manner
described here, it would be necessary to modify the Higgs sector, by adding
additional Higgs doublets for example.
\vskip .5cm
\noindent
[40] A massive mirror photon has quite unusual experimental
signatures. See, R. Foot, Mcgill preprint, hep-ph/9407331.
\vskip .5cm
\noindent
[41] In order for the mirror electron and mirror positrons
to annihilate sufficiently so that their relic abundances
do not overclose the universe, a lower bound (of about $10^{-5}$)
on the $U(1)$ kinetic mixing parameter can be calculated.
\vskip .5cm
\noindent
[42] It is not universally accepted that the standard big bang
model is the correct model for the origin of the universe.
For an alternative model, see e.g.
F. Hoyle, G. Burbidge, and J. V.
Narlikar, Astrophys. J. 410, 437 (1993); Astron. Astrophys. 289, 729
(1994).
See also, H. Arp et al., Nature 346, 807 (1990).
\vskip .5cm
\noindent
[43] Ya. B. Zel'dovich, I. Yu. Kobzarev, and L. B. Okun, Sov. Phys.
JETP, 40, 1 (1975).

\end{document}